\documentclass[letterpaper,fleqn,usenatbib]{mnras}

\def\eg{{\it e.g.}}
\def\etal{{\it et al.}}
\def\etc{{\it etc.}}
\def\ie{{\it i.e.}}

\def\Msun{M$_\odot$}

\def\pmb#1{\setbox0=\hbox{$#1$}%
  \kern-0.25em\copy0\kern-\wd0
  \kern.05em\copy0\kern-\wd0
  \kern-0.025em\raise.0433em\box0}
\def\spmb#1{\setbox1=\hbox{${\scriptstyle #1}$}%
  \kern-0.25em\copy1\kern-\wd1
  \kern.05em\copy1\kern-\wd1
  \kern-0.025em\raise.0433em\box1}

\long\def\Ignore#1{\relax}
\usepackage{color}
\definecolor{red}{rgb}{0.7,0.1,0.1}

\usepackage{graphicx}	% Including figure files
\title[Stability of M33]{The global stability of M33: still a puzzle}

\author[Sellwood, Shen \& Li]
          {J. A. Sellwood,$^{1}$\thanks{E-mail:sellwood@as.arizona.edu}
{Juntai Shen,$^{2,3,4,5}$\thanks{E-mail: jtshen@sjtu.edu.cn}
and {Zhi Li$^6$}
}
\\
$^1$Steward Observatory, University of Arizona, 933 N Cherry Ave, Tucson AZ 85722, USA \\
$^2$Department of Astronomy, School of Physics and Astronomy, Shanghai Jiao Tong University, Shanghai 200240, China (present address)\\
$^3$Key Laboratory for Research in Galaxies and Cosmology, Shanghai Astronomical Observatory, Chinese Academy of Sciences, \\ 80 Nandan Road, Shanghai 200030, China\\
$^4$Newton Advanced Fellow of the Royal Society of UK\\
$^5$School of Astronomy and Space Sciences, University of Chinese Academy of Sciences, 19A Yuquan Road, Beijing 100049, China\\
$^6$Tsung-Dao Lee Institute, Shanghai Jiao Tong University, Shanghai 200240, China}

\pubyear{2018}

\begin{document}
\label{firstpage}
\pagerange{\pageref{firstpage}--\pageref{lastpage}}
\maketitle

\begin{abstract}
The inner disc of the local group galaxy M33 appears to be in settled
rotational balance, and near IR images reveal a mild, large-scale,
two-arm spiral pattern with no strong bar.  We have constructed
$N$-body models that match all the extensive observational data on the
kinematics and surface density of stars and gas in the inner part of
M33.  We find that currently favoured models are unstable to the
formation of a strong bar of semi-major axis $2 \la a_B \la 3\;$ kpc
within 1~Gyr, which changes the dynamical properties of the models to
become inconsistent with the current, apparently well-settled, state.
The formation of a bar is unaffected by how the gas component is
modelled, by increasing the mass of the nuclear star cluster, or by
making the dark matter halo counter-rotate, but it can be prevented by
either reducing the mass-to-light ratio of the stars to $\Upsilon_V
\sim 0.6$ or $\Upsilon_K \sim 0.23$ in solar units or by increasing
the random motions of the stars.  Also a shorter and weaker bar
results when the halo is rigid and unresponsive.  However, all three
near-stable models support multi-arm spirals, and not the observed
large-scale bi-symmetric spiral.  A two-arm spiral pattern could
perhaps be tidally induced, but such a model would require an
implausibly low mass disc to avoid a bar and there is no visible
culprit.  Thus the survival of the current state of this exceptionally
well-studied galaxy is not yet understood.  We also suspect that many
other unbarred galaxies present a similar puzzle.
\end{abstract}


%%%%%%%%%%%%%%%%% BODY OF PAPER %%%%%%%%%%%%%%%%%%

\section{Introduction}
\label{sec.intro}
It has long been known \citep{Ho71, Ka78} that many models of galaxy
discs are unstable to the formation of a strong bar, yet a significant
fraction of disc galaxies lack any trace of a bar, and another large
fraction have only weak bars \citep{SW93, Es00, MD07, Ma11}.  Many
reviews have addressed how and why bars form \citep{To81, SW93, BT08,
  Se13} and both \citet{BS16} and \citet{BW19} highlighted the problem
of accounting for the observed bar fraction.  As these questions raise
a host of both theoretical and observational issues related to normal
mode analysis, halo responsiveness, bulges, non-linear behaviour at
resonances, tidal interactions, bar destruction, the role of gas,
galaxy formation and disc assembly, the evolution of the bar fraction
over cosmic time, secular changes, \etc, we will not review all this
material here and simply refer the reader to these articles.

Our puropse in this paper is to focus on one simply stated question:
can we understand why the Local Group galaxy M33, also known as NGC
598 or the Triangulum galaxy, does not possess a strong bar?  Recent
observational studies, most notably that by \citet[][hereafter
  C14]{Co14}, have provided a quite detailed level of information
about the distribution of stars, gas, and dark matter in the inner
parts of this galaxy, and some kinematic information about stellar
velocities near the disc centre.  Our objective is to create and
evolve a dynamical model that is as close a match as possible to the
observed constraints and to determine whether the model does or does
not form a strong bar in a short time period.  In posing this
question, we deliberately set aside questions of how the galaxy came
to be in its observed state, and assume only that the inner galaxy is
reasonably settled and is not presently undergoing rapid changes to
its internal structure.

The outer gas disc of M33 has long been known to be strongly warped
\citep{Ro76, Co14, Ka17} and the extended stellar disc is also
similarly warped \citep{Le13}, but to a slightly lesser extent.  This
feature has been attributed to a relatively recent passage of M33 past
its much larger neighbour M31 as modeled, for example, by
\citet{SLSA}.  However, the most recent proper motion measurements
\citep{vdM18} suggest that M33 has yet to pass close to M31, which if
confirmed would require a different origin for the misalignment of the
outer disc.

But whatever the cause of the warp, the careful analyses of the HI
emission by C14 and of the excited gas by \citet{Ka15} find the
kinematics of the gas inside a radius of $\la8\;$kpc of the nucleus of
M33 to be characterized by a well-ordered elliptical flow pattern,
that suggests a flat, rotationally-supported, nearly axisymmetric
disc, observed at an inclination to our line of sight.

\citet{HS80} identified a large number of spiral arms in M33 from B-
and V-band photometry, although many are just spurs and fragments.
Some of these features are also weakly visible in the outer parts of
the {\it WISE}~1 3.4~$\mu$m image presented by \citet{Ka15}.  These
last authors also fit multiple arms to their H$\alpha$ image.
However, near IR photometry at J, H and K \citep{RV94} and the 2MASS
image \citep{Ja03} reveal a regular 2-arm spiral pattern, with a third
weaker arm, and these features also dominate the inner disc in the
3.4~$\mu$m image.  We conclude that many of the arms counted by
\citet{HS80} reflect star-forming features that are bright in visible
bands due to young stars and begin to show up again at longer
wavelengths due to warm dust, but the old stellar disc, which is most
easily seen at wavelengths of a few $\mu$m, supports a more regular
bi-symmetric density wave.

\citet{RV94} note an apparent weak, inner bar, but were uncertain that
it was not just the inward extension of the spiral arms.  \citet{CW07}
reported mild non-circular motion in the inner 200~pc, which they
argued was indicative of a short, weak bar.  The analysis of the lower
spatial resolution HI observations presented in the appendix of C14
suggests that these mild non-axisymmetric features do not strongly
distort the circular flow pattern in the neutral gas.  This conclusion
was reinforced by \citet{Ka15}, who obtained small velocity residuals
from their measurements of the H$\alpha$ line, shown in their
Figure~11, after subtracting an axisymmetric flow pattern.

We conclude that this evidence supports our assumption that the inner
disc of M33, which contains most of the galaxy's stars and gas, is
indeed reasonably settled, lacks a strong bar, and is not presently
undergoing changes in its internal structure on an orbital time-scale,
which is $\sim400\;$Myr at $R=8\;$kpc.

Despite the abundance of data on M33, there have been rather few
attempts to model its internal structure in detail, although the warp
has attracted more attention \citep[see][and references
  therein]{SLSA}.  Both \citet{RK12} and \citet{Do18} focused on the
``flocculent'' spirals that are prominent in visual images
\citep{HS80}, which they model with hydrodynamics.  \citet{RK12}
compared different numerical recipes for stellar feedback over short
periods of evolution in a disc model embedded in a rigid halo that
approximated M33.  The simulations by \citet{Do18} had a dynamically
cool and heavy stellar disc with no halo in some cases, but their
simulations also covered a brief period and none continued the
evolution beyond 1~Gyr.  Here we report simulations of longer
duration, most of which form strong bars.

\begin{figure}
\includegraphics[width=.9\hsize,angle=0]{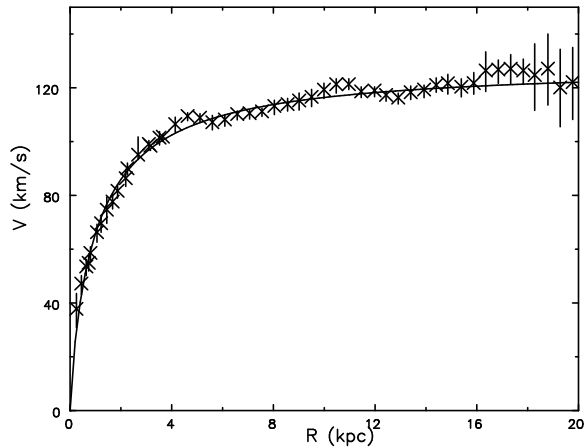}
% /home/sellwood/M33/vfit.s
\caption{The fitted rotation curve of M33.  The data points with error
  bars are from C14 and the smooth curve is our adopted fit
  (eq.~\ref{eq.rotcur}).}
\label{fig.rotcur}
\end{figure}

\section{Modelling M33}
We adopt the standard estimated distance to M33 of $840\;$kpc
\citep{Fr01, Ka15}, so that $1\arcsec$ projects to $\sim 4\;$pc, and
inclination of the inner disc of $i=52^\circ$ \citep{WWB, CS00}.

\subsection{Distribution of mass}
\label{sec.model}
C14 derived the deprojected rotation curve of M33 from Doppler
measurements of the 21~cm line of neutral hydrogen, which is in
excellent agreement with that obtained by \citet[][their
  Fig.~18]{Ka15} from a Fabry-P\'erot data cube of the H$\alpha$
emission line of excited hydrogen.  The radial variation of the
circular speed, with error bars, is reproduced here in
Fig.~\ref{fig.rotcur} from Figure~12 of C14, and takes account of the
pronounced warp in the disc of M33, which begins at $R \ga 8\;$kpc.
\Ignore{We have neglected their last few points for $R>20\;$kpc, which
  hint at a possible fractional decrease in circular speed, but the
  large uncertainties make them consistent with no decrease.}  The
solid curve indicates our least-squares fit to these data, which has
the following form
\begin{equation}
V(R) = A \left({R \over R + c}\right)^\alpha,
\label{eq.rotcur}
\end{equation}
where the values of the fitted parameters are: $A=128.42\;$km/s, $c =
0.80\;$kpc, and $\alpha =1.31$.  This form forces the curve to start
from the origin, which is not fully supported by the input data.
Although \citet{Ka15} obtain a gentle rise from a fit to their 2D
velocity map obtained by from Fabry-P\'erot observation of the
H$\alpha$ line, \citet{CW07} report a velocity jump across the nucleus
of $\sim 50\;$km~s$^{-1}$ between the approaching and receding sides
in their major axis slit position \citep[see also][]{Ru87}.  An
unresolved rise of the deprojected circular speed to $\sim
40\;$km~s$^{-1}$ would be consistent with the dense nuclear star
cluster, which has an estimated mass of $\sim
10^6\;$\Msun\ \citep{KM93, Ko10}.  Aside from this possible velocity
jump, which we will return to, we consider the smooth curve to be an
adequate fit to the data.

\begin{figure}
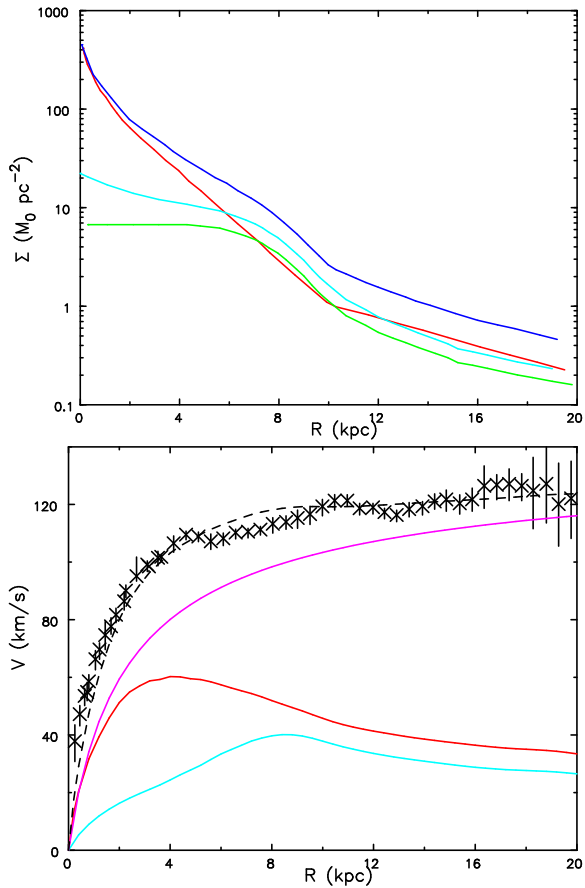

\includegraphics[width=.9\hsize,angle=0]{M33-Sigma.ps}
% /home/sellwood/M33/plot_data.s
\includegraphics[width=.9\hsize,angle=0]{M33-comps.ps}
% /home/sellwood/M33/get_halo.s
\caption{Top panel: The surface density profiles in \Msun$\; {\rm
    pc}^{-2}$, of the stars (red), HI (green), total gas (cyan), and
  total baryonic mass (blue).  Data are taken from the smoothed curves
  in C14, plus the molecular gas contribution from their eq.~(1).
  Bottom panel: rotation curves computed in the disc mid-plane from
  the disc stars (red) and the total gas (cyan), while the magenta
  line shows that from the fitted spherical dark matter halo.  The
  dashed black line shows the combined rotation curve of the three
  mass components, and the data points are reproduced from
  Fig.~\ref{fig.rotcur}.}
\label{fig.vcomps}
\end{figure}

C14 combined {\it BVIgi} photometry, SED fitting based on a revised
\citet{BC03} method, and the ``Padova 1994'' stellar evolutionary
library to create stellar mass maps for each pixel.  The black points
in their Figure 11, show azimuthally averaged values of the
mass-to-light ratio $\Upsilon$ in solar units of $1.1 \leq \Upsilon_V
\leq 1.3$ for $R>1.4\;$kpc, rising inside this radius to $\Upsilon_V
\simeq 1.5 \pm 0.2$.  C14 also find slightly higher values for
$\Upsilon_V$ when they use only the BVI colours.

\citet{Ka15} present surface brightness profiles derived from {\it
  Spitzer} IRAC 3.6~$\mu$m and {\it WISE}~1 3.4~$\mu$m images, and fit
an exponential disc with a length scale of 1.7~kpc with a small bulge
that accounts for 3\% of the total IR light.  \citet{Ka15} adopted
$\Upsilon_K = 0.72$ for the disc, giving a total stellar mass of $7.6
\times 10^9\;$\Msun \citep{Ka17}, which is much heavier than the
heavier disc model of C14, but these authors preferred a lower
$\Upsilon_K=0.52$ in their later paper, which yields a similar stellar
mass to that of the heavier disc proposed by C14.

Since heavier discs are more difficult to stabilize, we have adopted
the more conservative lower disc surface density of the two models
fitted by C14, which is presented in their Figure~10.  They
extrapolate the surface density from $R<6\;$kpc with two exponential
functions, adopting a slope change at $R=10\;$kpc, to give a total
mass for the stellar disc integrated to $R=20\;$kpc of $4.84 \times
10^9\;$\Msun.  In the same figure, they also give the HI surface
density they derived from the intensity of the 21~cm line, which they
correct for helium content to yield a smoothed neutral gas density
profile.  The upper panel of Fig.~\ref{fig.vcomps} shows the fitted
surface density profiles of the various baryonic components, which
includes the helium correction and the molecular contribution given by
eq.~(1) of C14.  The total mass of the gas disc integrated to
$R=20\;$kpc is $2.65 \times 10^9\;$\Msun.

We computed the central attractions of the stellar and total gaseous
discs separately by the elliptic integral method \citep{BT87}, which
takes account of the flattening of the disc to yield stronger central
attraction in the mid-plane and manifests features associated with the
changes in the slopes of the surface density.  We subtracted the
attractions of the disc stars and gas from the fitted circular speed
(eq.~\ref{eq.rotcur}), and attribute the remaining attraction to a
spherical dark halo in the usual way.  We adopt the cored halo density
profile
\begin{equation}
\rho_h(r) = {\rho_0 b^2\over (r + b)^2},
\label{eq.haldens}
\end{equation}
where the central density $\rho_0 = 0.212\;$\Msun\ pc$^{-3}$ and core
radius $b= 1.3\;$kpc.  The sums of the attractions of the two disc
components together with that of the halo yield our model rotation
curve, given by the dashed line, which is in satisfactory agreement
with the observed data.  Our non-standard halo density profile
(eq.~\ref{eq.haldens}) differs from both the NFW function \citep{NFW}
adopted by C14 and \citet{HW15} and the common pseudo-isothermal
function also used by \citet{Ka15}.\footnote{The denominator is
  $(r+b)^2$ and not $r^2 + b^2$.}

\subsection{Distribution functions}
We wish to realize an axisymmetric, equilibrium $N$-body model that
resembles the current properties of M33 as closely as possible.  In
order to achieve this, we must select particles from distribution
functions (DFs) for each component that would yield the desired
density profiles, together with any velocity dispersion constraints,
in the adopted total potential.

\subsubsection{Disc components}
As we are interested in the bar stability of the M33 disc, we limit
the radial extent of both the stellar and gaseous disc components by
applying an outer density taper that reduces their surface density
from the full observed value at $R=8\;$kpc to zero at $R=10\;$kpc
using a smooth cubic function.  The adopted truncation, which discards
almost 11\% of the stellar mass and 22\% of the gas mass, limits our
focus to the roughly co-planar inner part of the disc and neglects the
low surface density outer part, which is strongly warped and
halo-dominated.

\citet{KM93} used the Fourier quotient method to obtain a
line-of-sight velocity dispersion from the Ca II triplet absorption
line, finding values between 25 -- 30~km~s$^{-1}$ over the inner
1\arcsec\ of M33.  \citet{CW07} measured mostly emission line
velocities in multiple long slits at a variety of position angles.
However, they also estimated the stellar velocity dispersion ``along
the major axis of the inner disc'', again using the Ca II triplet,
finding values in the range 28 -- 35~km~s$^{-1}$, in agreement with
the other study.  Neither paper attempted to follow the outward radial
variation of the velocity dispersion, probably because of decreasing
surface brightness and/or instrumental spectral resolution limits.

We therefore adopt a uniform $Q=1.5$ for the stellar disc, where
\begin{equation}
  Q = {\sigma_R \over \sigma_{R,\rm crit}} \qquad\hbox{with}\qquad
  \sigma_{R,\rm crit} = {3.36 G\Sigma \over \kappa},
\end{equation}
\citep{To64}.  Here $\sigma_R$ is the Gaussian width of the radial
velocity component of the particles, and $\kappa(R)$ is the epicyclic
frequency.  Adopting a constant $Q$ yields a velocity dispersion in
both the radial and azimuthal components that decreases outwards from
a central value of $\sigma_R \sim 35\;$km~s$^{-1}$.  Our model also
has a central value of the vertical velocity dispersion of $\sigma_z
\sim 21\;$km~s$^{-1}$, which is constructed as described below.  While
the true axis ratio of the velocity ellipsoid in M33 is unknown, a
flattening of $\sim 1.7:1$ may be reasonable.  Combining the in-plane
and vertical components at an adopted inclination of $52^\circ$ yields
a line-of-sight dispersion of $\sim 31\;$km~s$^{-1}$, which is
consistent with reported values.  We determine the appropriate
in-plane DF by the method given by \citet{Sh69} in the numerically
determined attraction of the thickened discs and halo mass
distributions.

We also model the gaseous disc with collisionless particles in some
of our simulations.  In these cases, we adopt $Q=0.1$ and select
velocities for the particles using the Jeans equations \citep{BT08}
which are adequate when $\sigma_r \ll V_c$.

We adopt a Gaussian function with a spread of 100~pc in the
$z$-direction to create the volume density profiles of both the star
and gas discs and assign velocities using the vertical Jeans equation
at each radius in the numerically determined gravitational attraction.

These procedures allow us to realize the discs with particles that
result in satisfactory equilibria at the outset.  The radial variation
of the velocity dispersion in the stellar disc is shown in the lower
panel of Fig.~\ref{fig.modelrc}.

\subsubsection{Halo component}
\citet{HBWK} demonstrated that a satisfactory equilibrium halo DF can
be determined by Eddington inversion \citep{BT08} in the potential of
a composite disc-halo mass distribution by neglecting the flatness of
the disc \ie, the attraction of the disc should be $GM_d(r)/r^2$,
where $M_d(r)$ is the disc mass enclosed within a sphere of radius
$r$; this spherical average is generally weaker than the attraction in
the disc mid-plane.  They showed that the mild flattening of the final
potential has a negligible effect on the equilirium of the halo, as we
have also found in our own work.  We therefore use Eddington's formula
to derive an isotropic DF for the halo of M33 in the presence of the
disc components.

Since our adopted halo density (eq.~\ref{eq.haldens}) yields an
enclosed mass that asymptotically increases linearly with radius, we
limit its extent by setting the density to zero for $r > r_{\rm
  trunc} = 40\;$kpc.  In previous work, we have found it acceptable to
limit the radial extent of an infinite live halo having a steeply
declining density profile simply by computing the DF assuming an
infinite radial extent and then discarding from the DF any particles
having $E > \Phi(r_{\rm trunc})$, where $\Phi(r)$ is the gravitational
potential of the total mass distribution.  This bound eliminates all
particles having sufficient energy that even part of their orbits
extend beyond $r_{\rm trunc}$, thereby creating a smooth density
decrease to $\rho = 0$ at $r = r_{\rm trunc}$.  Even though the
resulting particle density falls below that required to maintain the
central attraction in the outer parts, we have generally found that
the consequential slight disequilibrium of the outer halo led to
negligible adjustments to the outer density profile as the model
evolved.

However, this was not the case with our current halo model, whose
density decreases with radius only as $r^{-2}$.  An isotropic DF
within the shallower potential well of this halo includes many orbits
that have a wide radial range, and our previously adopted truncation
rule resulted in a density of active particles that fell far below the
expected density even inside $r < r_{\rm trunc}/2$, with the result
that the halo model began to expand, reducing the density of all but
the very inner part, as it relaxed towards equilbrium.  Inclusion of
the boundary term from the inversion formula \citet{BT08}, which we
had at first neglected, made only a minor improvement.

We therefore have revised the halo density profile to taper to zero at
$r_{\rm trunc}$ as
\begin{equation}
\rho_h(r) = {\rho_0 b^2\over (r + b)^2} \cases{ \cos\left({\pi r \over 2r_{\rm trunc}}\right) & $r<r_{\rm trunc}$ \cr 0 & otherwise, \cr}
\label{eq.rhaldens}
\end{equation}
and compute the DF in the total potential of the discs plus this
revised halo by Eddington inversion which, with the inclusion of the
boundary term, results in an equilibrium halo.  The cosine taper
shaves mass from the halo at all radii, but it has a negligible effect
on the halo mass interior to 10~kpc, which is the radial extent of the
discs, and the circular speed is decreased by just 5\%, or
7~km~s$^{-1}$, at $r=20\;$kpc.

\subsection{Combined model}
\begin{figure}
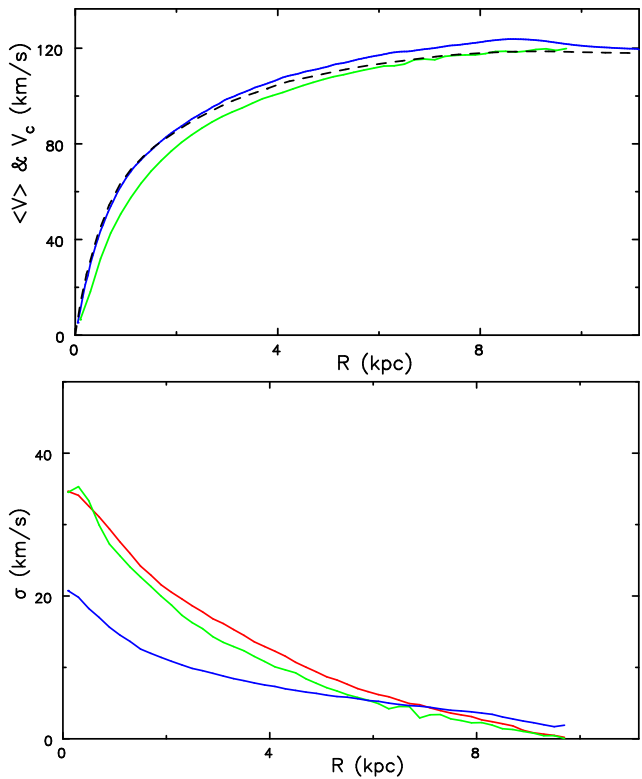

\includegraphics[width=.99\hsize,angle=0]{modelrc.ps}
\includegraphics[width=.99\hsize,angle=0]{veldisp.ps}
% /home/sellwood/M33/modelrc.s /home/sellwood/M33/veldisp.s
\caption{Upper panel: The rotation curve of our initial model.  The
  blue curve shows the circular speed in the disc plane computed from
  the attraction all three mass components, the green curve indicates
  the mean orbital speed of the star particles, while the dashed line
  is drawn from eq.~(\ref{eq.rotcur}).  Lower panel: The radial
  variation of the dispersions of the radial (red), azimuthal (green),
  and vertical (blue) velocity components in the initial stellar
  disc.}
\label{fig.modelrc}
\end{figure}

The rotation curve in the mid-plane of the combined three-component
model, which takes account of disc thickness and gravity softening, is
given by the blue curve in the top panel of Fig.~\ref{fig.modelrc}; it
is in reasonable agreement with the dashed line that is drawn using
eq.~(\ref{eq.rotcur}).  The bump near $R=8\;$kpc is caused by our
truncation of the disc components at that radius \citep[see
  \eg][]{Ca83}.  The mean orbital speed of the star particles, green
curve, is lower because of asymmetric drift.  The virial ratio of the
complete model is $T/|W| \simeq 0.495$, and no significant adjustment
to the mass or velocity distribution of any component occurs in the
initial evolution.

\begin{figure*}
\includegraphics[width=.99\hsize,angle=0]{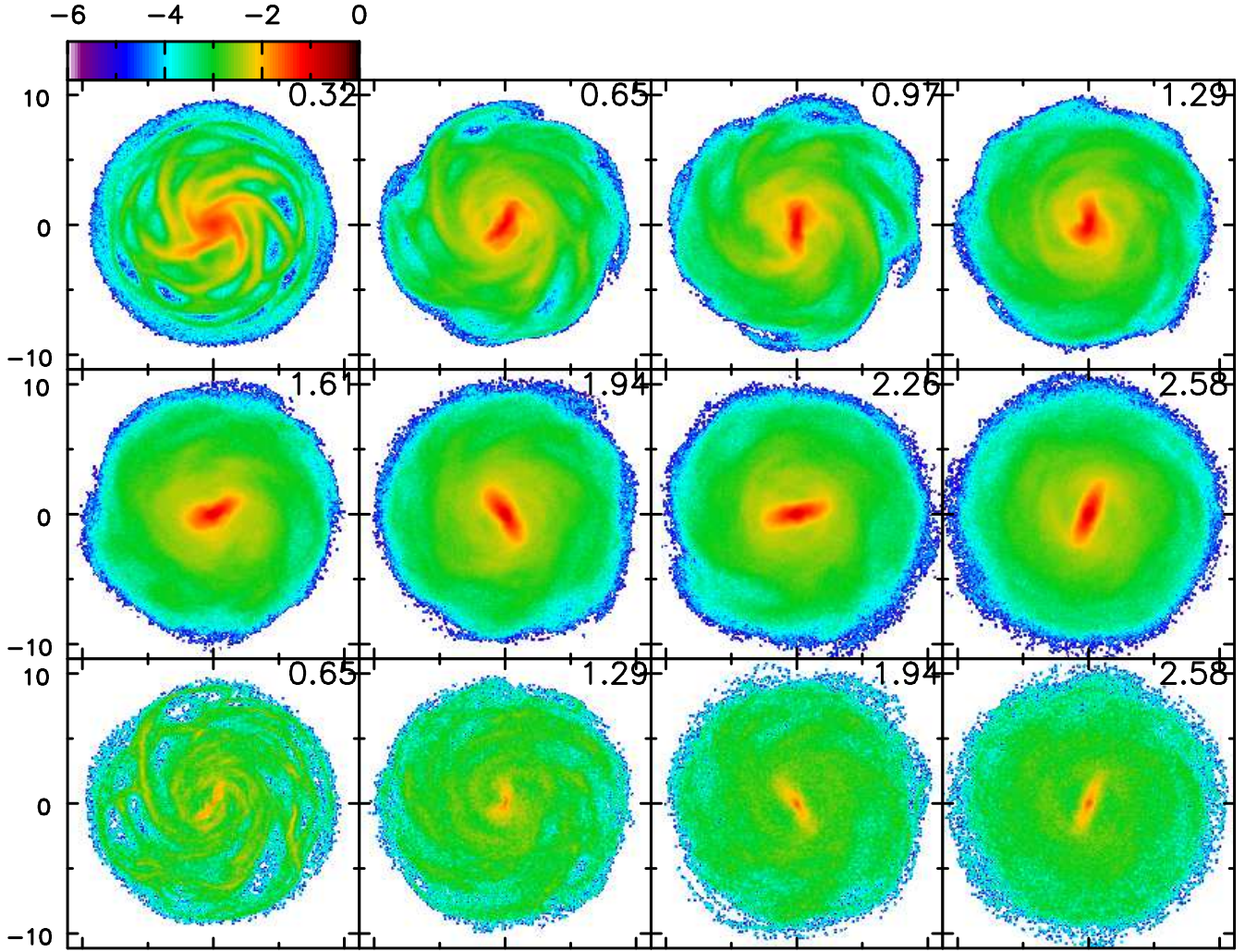}
% /home/sellwood/M33/denevol.s
\caption{The evolution of our fiducial simulation.  The first 16
  panels show the projected surface density of the stellar disc, the
  botton row displays the evolution of the cool component on a coarser
  time interval.  Times in the top right corner of each panel are in
  Gyr, the axes are marked in kpc, and the colour scale shows the
  logarithm relative to the maximum in each panel.}
\label{fig.denevol}
\end{figure*}

\section{Simulations}
\label{sec.sims}
Both the stellar disc and halo were each represented by 1 million
particles in our fiducial simulation.  These collisionless particles
were drawn from the DFs described above using the smooth procedure
outlined in the appendix of \citet{DS00}.  The gaseous disc was
represented by $100\,000$ particles, chosen at random from the
truncated gas density profile.

\subsection{Collisionless particle only simulations}
\label{sec.galaxy}
We computed the evolution of this three component model using the
hybrid option of the {\tt GALAXY} code \citep{Se14}.

\subsubsection{Numerical parameters}
The gravitational attraction of the two disc components was computed
using a 3D cylindrical polar grid having 330 rings, 512 spokes and 75
planes.  The outer edge of the grid was at 11.13~kpc, the vertical
spacing of the grid planes was 50~pc, so that it extended to $\pm
1.85\;$kpc from the disc mid-plane.  We softened gravity using the
cubic-spline kernel recommended by \citet{Mo92}, which yields the full
Newtonian attraction at distances $\geq 2\epsilon$, where the
softening length $\epsilon = 50\;$pc.  We further smooth away
small-scale density fluctuations by discarding sectoral harmonics
$m>8$ in the solution for the gravitational field.

The field of the halo particles was computed using a spherical grid
having 200 shells and extending to 42~kpc, and we used a surface
harmonic expansion to compute aspherical terms of the field on each
shell, up to and including the $l=4$ terms.  Naturally, the mutual
attraction of the disc and halo particles was included.

We employed a basic time step of $67\,000\;$years, which was increased
by two factors of two at spherical boundaries of radii 1~kpc and
2.5~kpc.  As usual, we ran comparison models in which all numerical
parameters were varied within reasonable ranges, which generally
caused little change, except that a ten-fold increase in the number of
particles in each component caused bar formation to be slightly
delayed, since the seed amplitude of the instability was lower, while
increasing the softening length to 200~pc also resulted in slower bar
formation, but the ultimate outcome was unchanged.

\subsubsection{Basic result}
The evolution of our fiducial model is shown in
Fig.~\ref{fig.denevol}.  The initially axisymmetric disc forms
multi-arm spirals at first, in agreement with the behaviour reported
by \citet{Do18} in their short simulations.  However, a pronounced bar
becomes established before 1~Gyr of evolution, which settles and
persists to the end of the simulation.  The bar forms through the
standard global bar instability mechanism \citep{To81, BT08} and has
no inner Lindblad resonance.  The semi-major axis $a_B \ga 2.5\;$kpc
when it first forms and $a_B \sim 3\;$kpc by the end.  The study by
\citet{SLSA} was focused on the tidal interaction with M31, but these
authors reported a simulation of M33 in isolation that also formed a
bar in the inner 3~kpc of the disc.  Their simulation included gas
dynamics, star formation, and feedback but they gave few details about
the bar.

Our fiducial simulation includes a disc component having a much lower
initial velocity dispersion and the density profile of the total gas.
Since we also modelled this component using collisionless dynamics, it
is hardly surprsing that its evolution, shown in the bottom four
panels of Fig.~\ref{fig.denevol}, follows that of the more massive
stellar disc.  We report simulations in \S\ref{sec.gizmo} in which
this component is treated more like a gas.

The instability we find that forms a strong bar on the timescale of a
few orbit periods is a surprising result, since it seems likely that
M33 is stable today.  To argue otherwise, is to assume that we see M33
at a special time, when something has happened recently to make its
inner disc unstable, while it was stable $<1\;$Gyr ago.  It might be
argued that the triggering event was a possible tidal interaction with
M31; this interaction may have created the warp, as modelled by
\citet{SLSA}.  But there is no evidence for a disturbed flow pattern
in the distribution and kinematics of the stars and gas of the inner
disc, and this is the region that would need to have been changed
recently to have tipped the disc from stability into the unstable
state we find.

\begin{figure}
\includegraphics[width=\hsize,angle=0]{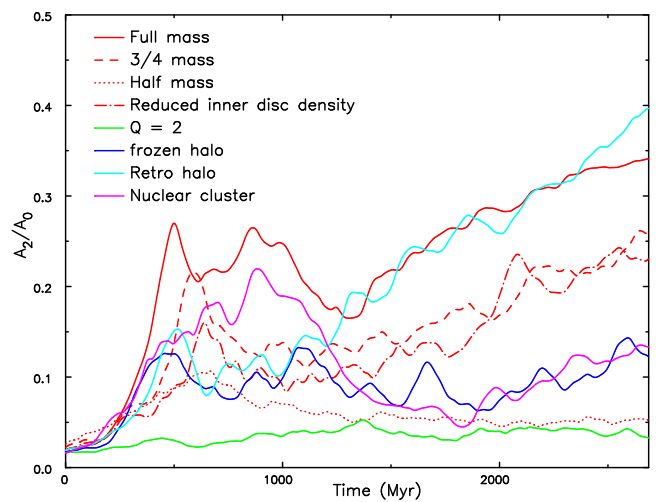}
% /home/sellwood/M33/ampevol_all.s
\caption{The time evolution of the bar amplitude in eight
  simulations. The ordinate is the relative amplitude of the $m=2$
  component determined over the radial range $0.5<R<3\;$kpc.}
\label{fig.ampevol}
\end{figure}

\subsubsection{How could a bar be avoided?}
\label{sec.avoid}
We therefore discount the possibility of a recent change to the
stability of the inner disc that has not yet had time to manifest
itself.  Instead we address the question of how must the properties we
have adopted for M33 be changed in order that the disc remains stable.
We have run additional simulations to examine all of the following
possibilities, and compare, in Fig.~\ref{fig.ampevol}, the time
evolution of the bar amplitude in each case, with that in our fiducial
model (solid red curve).  We discuss the implications of these results
in the next section.

\begin{enumerate}
\item \citet{At02} and others have shown that bars grow more rapidly
  in responsive halos than in rigid ones.  We have therefore
  experimented with replacing the halo by the fixed central attraction
  that would arise if the halo particles were replaced by a rigid mass
  distribution.  As expected, this change substantially reduced the
  growth rate of the bar, as shown by the blue curve in
  Fig.~\ref{fig.ampevol}, but a weak bar of length $\sim 2\;$kpc was
  present by the end of the simulation.

\item \citet{SN13} found that bars grew yet more rapidly when the halo
  was given some angular momentum in the same sense as the disc, and
  conversely growth was slowed slightly when the halos rotated in a
  retrograde sense.  We have therefore experimented with adding a
  large amount of angular momentum to the halo to make it
  counter-rotate against the disc, to almost the maximum extent
  possible by reversing the sign of the $z$-component angular momentum
  of most of the halo particles having $L_z>0$ initially.  Although
  this change slowed the initial growth of the bar, as indicated by
  the cyan curve in Fig.~\ref{fig.ampevol}, the bar amplitude rose
  steadily until it exceeded that in the isotropic halo (red curve).

\item Reducing the disc mass, while increasing the halo mass to
  preseve the same total rotation curve has historically been regarded
  as the favoured way suppress the bar instability \citep{OP73, ELN,
    Ch95}.  We report two experiments in which we reduced the stellar
  disc mass to 75\% and to 50\% of the values we adopted from C14.
  (The mass of the gas component was unchanged.)  In both cases we
  fitted a new halo density profile to compensate for the reduced disc
  mass and derived a new isotropic DF for it by Eddington inversion.
  The evolution of the bar amplitude in these cases is shown in
  Fig.~\ref{fig.ampevol}.  The dashed red line (75\% disc mass)
  indicates that a slightly shorter ($a_B \simeq 2\;$kpc), but still
  pronounced, bar developed in this case, while no significant bar
  developed when the disc mass was halved (dotted red line).

\item \citet{BS16} were able to stabilize some of their galaxy models
  by reducing the surface density of the inner disc only.
  Accordingly, we tried tapering the inner disc density from the full
  density at $R=2\;$kpc to half the central density at $R=0$.  The
  resulting reduced attraction of the inner disc allows an
  approximately four-fold increase in $\rho_0$, the central density of
  the halo, and a smaller core radius $b$.  The dot-dashed curve in
  Fig.~\ref{fig.ampevol} shows that a bar still forms; the final bar
  is somewhat shorter ($a_B \simeq 2\;$kpc) than in our fiducial run.

\item \citet{AS86} demonstrated that the bar instability could also be
  quelled by increasing $Q$.  The green curve in
  Fig.~\ref{fig.ampevol} shows that the amplitude of a possible
  bar-like feature remains very low when disc random motion is raised
  by 33\% to $Q=2$, and the initial thickness of the stellar disc was
  also doubled.

\item \citet{To81} argued, on the basis of linear perturbation theory,
  that the bar instability could be inhibited by a dense center
  \citep[see also][]{Se89, SE01, SE18}.  As noted above, we have so
  far ignored the nuclear star cluster in M33 whose properties were
  characterized by \citet{KM93}.  As the mass of this object is small
  \citep[][give $1.2\pm0.2\times 10^6~$M$_\odot$]{Ko10}, we thought it
  unlikely to affect the global dynamics of the disc, but we include
  it in an additional test for completeness.  Here we model the
  nuclear cluster as a rigid central Plummer sphere having a mass of
  $2.63 \times 10^7\;$\Msun\ and a core radius of 10~pc, which is over
  20 times the estimated mass and also more diffuse than the nuclear
  star cluster in M33, but causes the circular speed to rise to $\sim
  64\;$km~s$^{-1}$ at $R=11.4\;$pc or 2.85\arcsec.  This dense
  component required us to shorten the basic timestep by a factor of
  10.  As expected, the initial bar amplitude evolution, magenta curve
  in Fig.~\ref{fig.ampevol}, is little affected by this small extra
  mass.  Although the bar length $1.5 \la a_B \la 2\;$kpc does not
  change much, its amplitude is substantially weakened as the
  evolution continued.  The formation and subsequent weakening of the
  bar almost doubles the velocity spread of particles in the inner
  disc, making this mildly barred model at later times again
  inconsistent with the current state of M33.
\end{enumerate}

%%% GIZMO figures %%%%%
\begin{figure*}
\includegraphics[width=\hsize,angle=0,clip=]{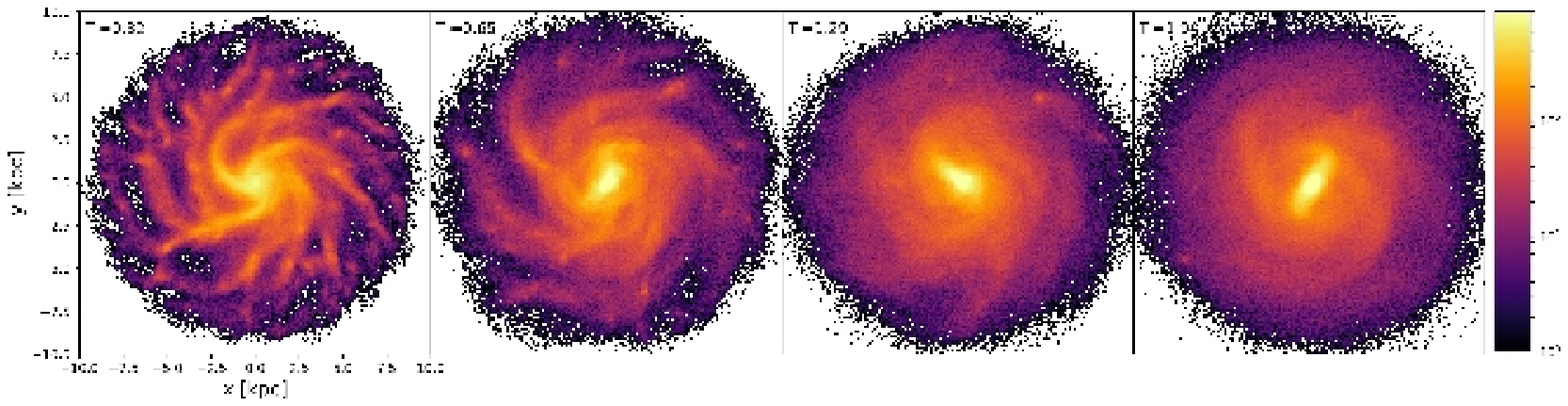}
\includegraphics[width=\hsize,angle=0,clip=]{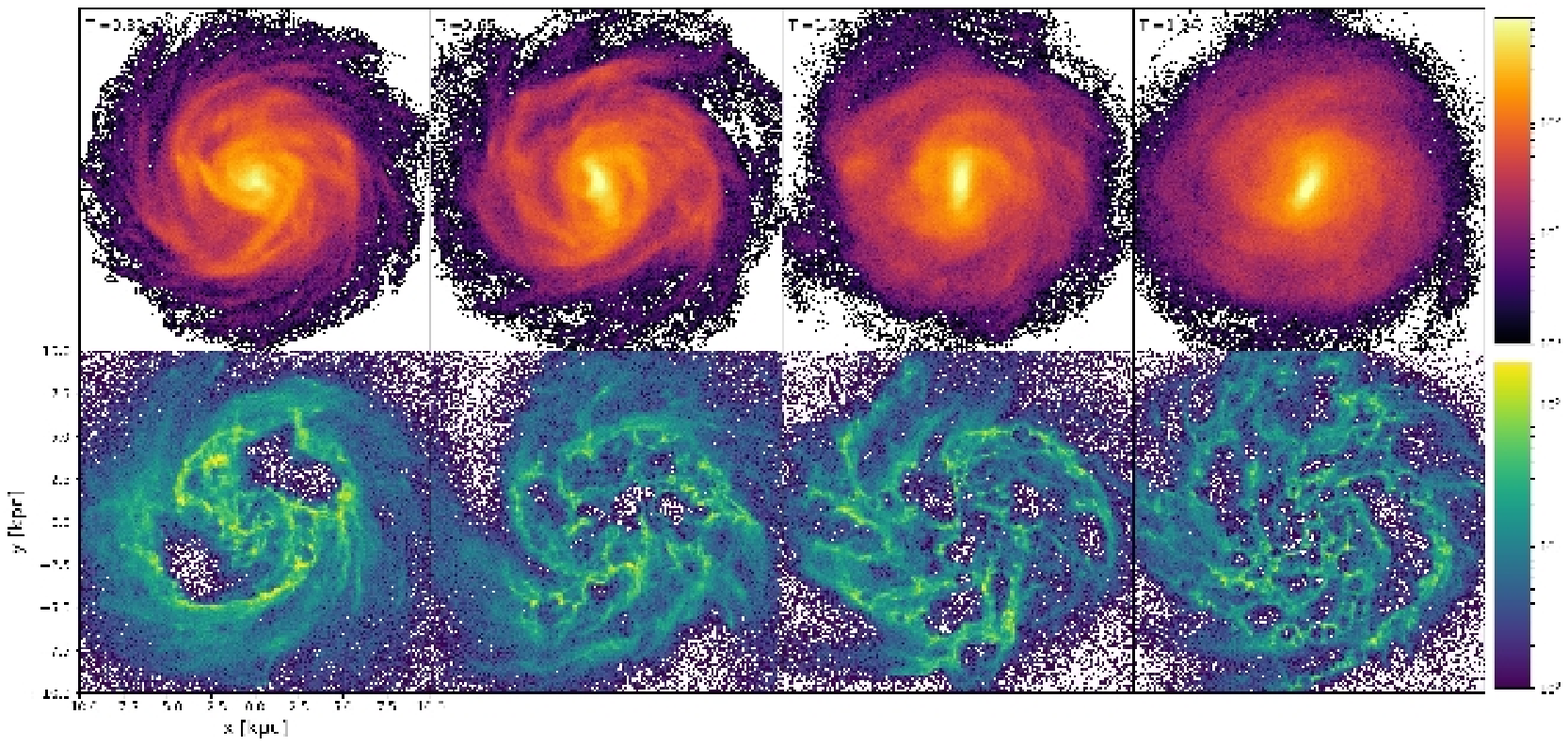}
\includegraphics[width=\hsize,angle=0,clip=]{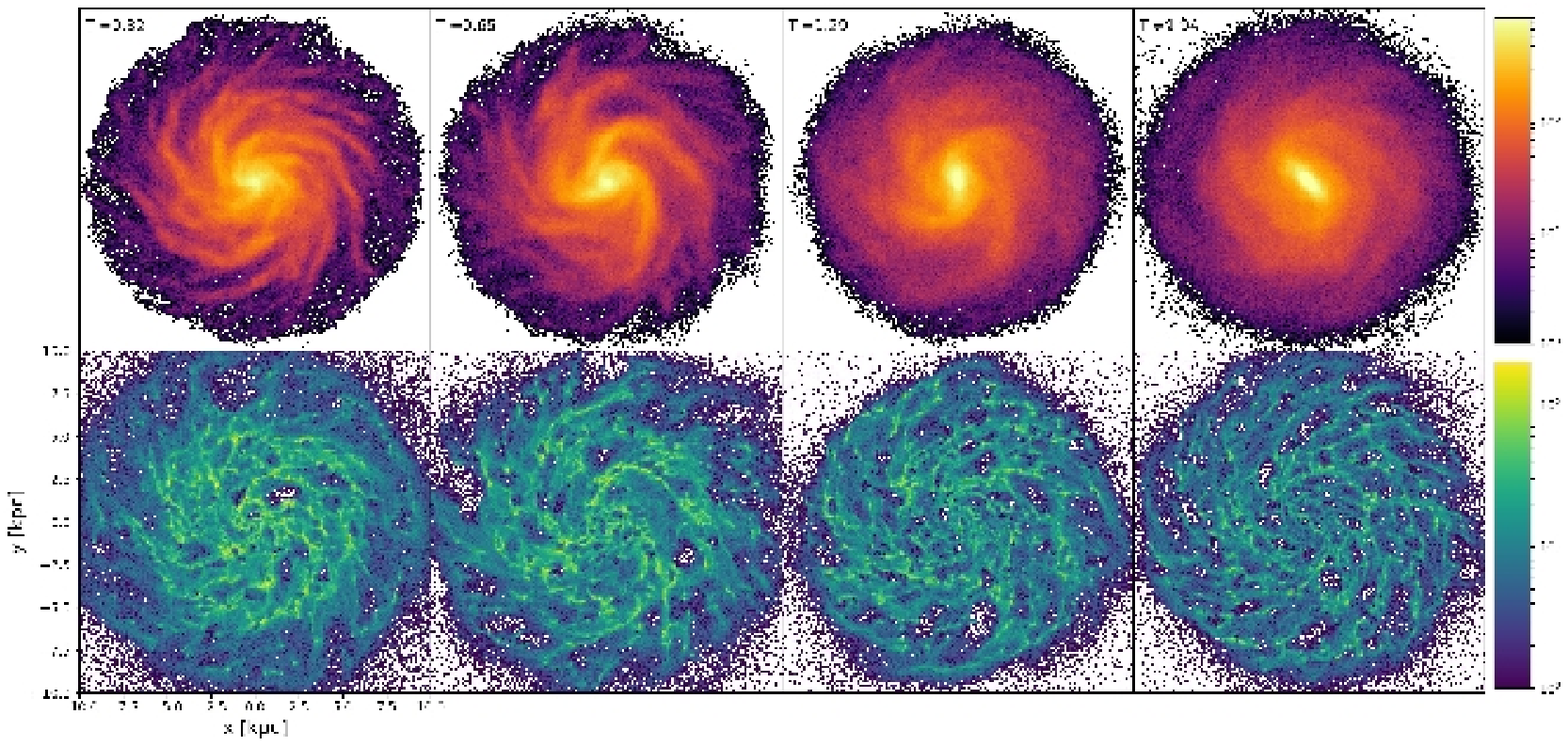}
\caption{The evolution of simulations using the {\tt GIZMO} code. Top
  row: The time evolution of the stellar surface density when the
  ``gas'' particles are also treated as collisionless, for comparison
  with our our fiducial model shown in Fig.~\ref{fig.denevol}.  Second
  two rows: The time evolution of the surface density of stars (above)
  and gas (below) in the simulation with mechanical feedback.  Bottom
  two rows: As in the middle two but for the case in which thermal
  feedback has been adopted instead of mechanical feedback.}
\label{fig.gizmo}
\end{figure*}

\subsection{Modelling the gas}
\label{sec.gizmo}
All the above models employed a second set of collisionless particles
for the gas component.  Here we present simulations using a different
code to determine whether the bar instability is altered by a
hydrodynamical treatment for this component, both with and without
star formation and feedback.  We use the publicly available {\tt
  GIZMO} code\footnote{We downloaded the public version at:
  \hfil\break {\tt
    http://www.tapir.caltech.edu/{\footnotesize$\sim$}phopkins/Site/GIZMO.html}}
\citep{Ho15} which is able to simulate both hydro- and stellar
dynamics.  The $N$-body part of {\tt GIZMO} is almost identical to
that in {\tt GADGET3}, and the code offers several modern
hydrodynamical solvers which share some common features with
moving-mesh codes such as {\tt Arepo} \citep{Sp10}.  We adopt the
recommended Meshless Finite-Mass method for the hydrodynamics.  The
minimum softening length for gas and star particles is 50~pc, while
that for dark matter particles is 100~pc.

In most of our simulations with {\tt GIZMO}, we employed five times as
many gas particles as we used in {\tt GALAXY}, but otherwise the
initial positions and velocities of star, gas, and dark matter
particles are the same as in our fiducial model.  The outcomes of tests
with gas physics that employed the smaller number of gas particles
were very similar to those with the larger number, but all the
hydrodynamical results we present here use the larger number of gas
particles ($N_{\rm gas}=5 \times 10^5$).  We also tried mimicking the
effect of the nuclear star cluster (NSC) in M33 by including a massive
particle of $10^6$\Msun\ (the mass of $\sim 250$ star particles) and a
softening length of $5\;$pc.  Since we found that including or
omitting the NSC had almost no effect on the bar instability, we here
present only results without a NSC.

The top row of Fig.~\ref{fig.gizmo} presents a check of the evolution
using the {\tt GIZMO} code with both disc components treated as
collisionless particles for comparison with our fiducial {\tt GALAXY}
model.  The agreement with the result shown in Fig.~\ref{fig.denevol}
is entirely satisfactory; the bar amplitude at $t\sim2.4\;$Gyr is
$A_2/A_0 \sim 0.3$ and the bar lengths are very similar in both codes.

We have experimented with different ways to model the gas component in
order to test their effect on the bar instability.  The second two
rows of Fig.~\ref{fig.gizmo} show the behaviour of a simulation with
gas that includes radiative heating and cooling by metal species, star
formation, and mechanical feedback from star formation \citep{SH03,
  KG11, HNM, Ho17}.  The result in the bottom two rows of
Fig.~\ref{fig.gizmo} shows the different behaviour when thermal,
instead of mechanical, feedback is used.  A strong bar forms in the
stellar component by $t\sim0.5\;$Gyr in both cases, and weakens only
slightly after $t\sim1.5\;$Gyr.  It is still very prominent in the
bottom two rows, where the bar amplitude is $A_2/A_0 \sim 0.2$ at
$t\sim 2\;$Gyr. The mechanical feedback creates larger ``holes'' in
the gas in (middle two rows) than happens with just thermal feedback
(bottom), in agreement with the findings of \citet{Do18}.

\begin{figure*}
\includegraphics[width=.9\hsize,angle=0]{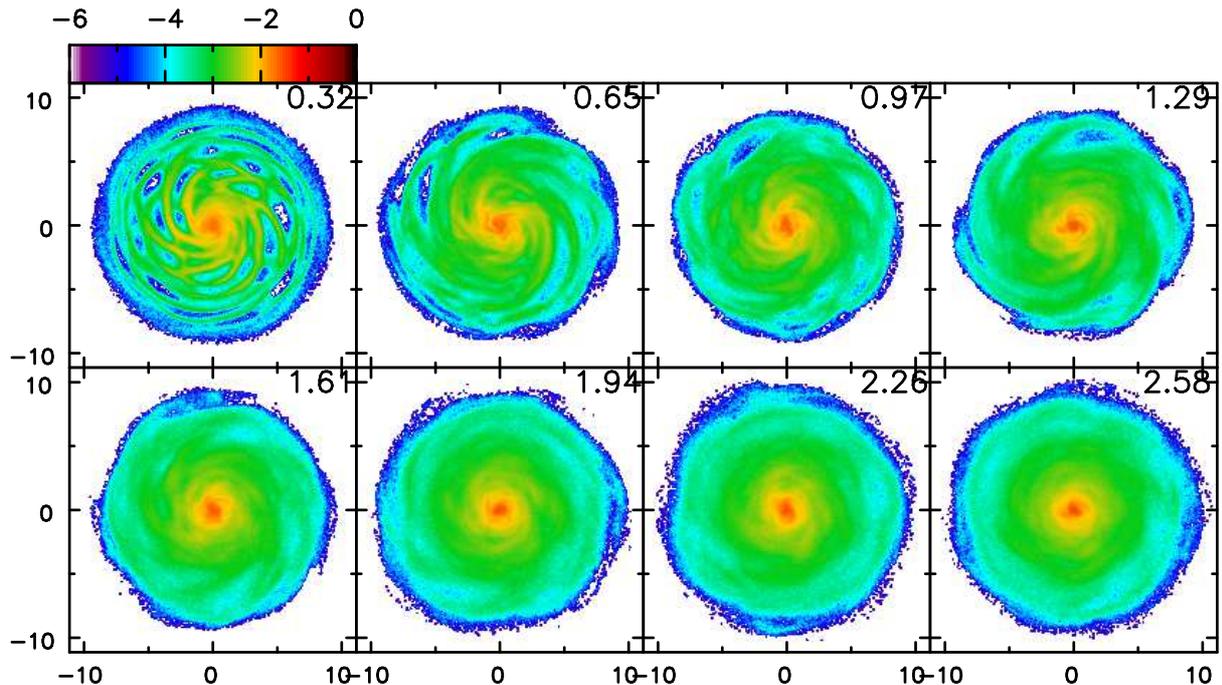}
% /home/sellwood/M33/light.s
\caption{The evolution of the stellar component when the stellar disc
  mass is halved from that in the fiducial model.  As in
  Fig.~\ref{fig.denevol}, times in the top right corner of each
  panel are in Gyr, the axes are marked in kpc, and the colour scale
  shows the logarithm relative to the maximum in each panel.}
\label{fig.light}
\end{figure*}

We have also tried other prescriptions to model the gas; in one case
treating it as an isothermal gas ($T=12\,000$K), and another as having
an adiabatic equation of state ($\gamma=5/3$) with cooling, but
without feedback in both cases.  We do not show the evolution of these
simulations, but note that a strong bar also formed in both and the
absence of feedback allowed a large fraction of the gas to be driven
into the centre.  The inflow in the adiabatic case was sufficient to
cause the bar to weaken at later times.

In summary, we find that a strong bar still forms independently of how
we choose to model the hydrodynamics of the gas component.

\section{Discussion}
\label{sec.discussion}
We have found that a strong bar forms in our stars-only models unless
(a) the stellar disc mass is halved, (b) the velocity dispersion and
disc thickness are increased, or (c) the responsive halo is replaced
by a rigid halo.  While a bar also formed in all our simulations with
gas, it subsequently dissolved in one case apparently due to gas
inflow to the center, as also happened in the {\tt GALAXY} simulation
with a central mass to represent an excessively massive nuclear star
cluster.  But the early formation of a bar confirmed that the {\it
  current state} of M33 is unstable and the subsequent evolution
altered the mass distribution and velocity dispersion to be
inconsistent with that currently observed.

\subsection{A rigid or rotating halo}
\label{sec.rigid}
Our results have confirmed the finding by \citet{At02}, \citet{SN13},
and \citet{BS16} that galaxy models that might be bar-stable in rigid
halos can be unstable when the same halo is composed of responsive
particles.  While it has long been known \citep{Se80, We85} that
strong bars can exchange angular momentum with a responsive halo,
\citet{Se16} showed that the bar instability in the disc could also
elicit a supporting response from mobile halo particles even in the
linear growth phase, when amplitudes are tiny.  Thus live haloes
exacerbate the bar stability problem \citep{At02}, as was confirmed by
our test reported in Fig.~\ref{fig.ampevol}.  Changing the halo DF
to include rotation \citep{SN13} or velocity anisotropy \citep{Se16}
has some effect on the growth rate of the bar, but freezing the halo
reduces it to a much greater extent.

Thus the stability criteria proposed by \citet{OP73}, \citet{ELN}, and
\citet{Ch95} do not apply in more realistic models.  \citet{Se16}
found that a disc in a live halo can still be stabilized by reducing
its mass, and estimated that the ratio of disc-to-halo attraction must
be reduced to between 30\% and 50\% of that required for a rigid halo.
Nevertheless, our disc model embedded in a rigid halo still formed a
weak bar with $a_B \simeq 2\;$kpc and having about 1/3 the amplitude
of that shown in Fig.~\ref{fig.denevol}, while the live halo model
with half the disc mass was stable, which is consistent with these
earlier results.

\subsection{Stability and disc mass}
\label{sec.halos}
C14 estimated a mean $\Upsilon_V \sim 1.2$ over the radial range $1.5
\la R \la 5\;$kpc from modeling surface brightness measurements in the
$BVIgi$ colour bands, and a slightly higher value from $BVI$
photometry alone.  Adopting the lower mass, we have found that the
disc forms a strong bar of semi-major axis $a_B \sim 3\;$kpc.  We also
obtained a strong, but slightly shorter, bar when we tried reducing
$\Upsilon_V$ to 75\% of their lower value, but we found no evidence
for a bar throughout the evolution in a model in which we halved the
disc mass from that in our fiducial model.  In this last case, our
model assumed that $\Upsilon_V < 1$ everywhere and $\Upsilon_V \sim
0.6$ over most of the disc.  \citet{Ka17} reported that their
originally favoured $\Upsilon_K=0.72$ \citep{Ka15} yielded a stellar
disc mass of $7.6 \times 10^9\;$\Msun, while we had to reduce the disc
mass to $2.42 \times 10^9\;$\Msun\ to obtain stability, which would
seem to imply $\Upsilon_K \sim 0.23$ for our stable disk model.  While
$\Upsilon$ values derived from photometric broadband colours can be
quite uncertain, because the fraction of mass in dim, low-mass stars
is not well constrained, $\Upsilon_V \sim 0.6$ and $\Upsilon_K \sim
0.23$ are lower than typically expected values \citep[\eg][]{Co13,
  SML18} and would imply an unusually top-heavy IMF in the disc of
M33.  We also note that this low value for $\Upsilon$ makes the
stellar mass within 20~kpc less even than the corresponding gas mass.

Furthermore, if it were the case that the presence of bar in a galaxy
is determined by halo dominance, then barred galaxies would have
systematically heavier discs than their unbarred cousins.  Such a
difference would be manifested in a systematic offset in the
Tully-Fisher relation, since unbarred galaxies of a given luminosity
would be predicted to have higher circular speed, which is not
observed \citep{MF96, Cour03}.  Indeed, \citet{Bo96} was unable to
find any systematic differences between barred and unbarred galaxies.

Tapering the surface density of the inner disc to half its value at
$R=0$, which \citet{BS16} reported could stabilize some models, did
not prevent a strong bar from forming in the case of M33.  As noted
above, this change effectively reduces $\Upsilon$ in the inner disc to
be below that of the outer disc, which seems the wrong way round.

\subsubsection{Spiral constraints on disc mass}
It should also be noted that all the spiral patterns in our simulation
of this low mass disc were multi-armed, as shown in
Fig.~\ref{fig.light} and there was little sign of the dominant $m=2$
pattern that stands out in near-IR images of M33 \citep{RV94, Ja03}.
The multiplicity of spiral arm patterns that develop through
gravitational instabilities in a stellar disc is determined by the
swing-amplification parameter $X$ \citep[defined by][]{To81}; vigorous
amplification occurs over the range $1.5 \leq X \leq 2.5$ when the
rotation curve is flat, and for smaller values when the rotation curve
rises.  Recall that $X = 2\pi R_c / (m\lambda_{\rm crit})$, which is
the ratio of the azimuthal wavelength of an $m$-armed spiral to the
characteristic scale for gravitational instabilities in discs
$\lambda_{\rm crit} = 4\pi^2G\Sigma / \kappa^2$ \citep{To64},
evaluated at $R=R_c$, the corotation radius of the spiral.  Thus the
preferred arm multiplicity $m$ in a family of models having a fixed
rotation curve rises as the inverse of the disc surface mass density
$\Sigma$ \citep{SC84, At87}.  The low-mass disc whose evolution is
shown in Fig.~\ref{fig.light} has $Xm \sim 6$ at $R=1\;$kpc, where the
rotation curve is still rising, and this quantity increases to $Xm
\sim 16$ at $R = 4\;$kpc, accounting for the $m \geq 4$ spirals that
were preferred in this model.  Thus the fact that the clearest spiral
pattern at near-IR wavelengths is bi-symmetric in M33 is an additional
argument against preventing bar formation by reducing the disc mass.
In fact, the great majority of spiral galaxies, both with and without
bars, have two spiral arms \citep{Davi12, Hart16, Yu18}, which is an
indicator of a heavy disc \citep{SC84, At87}.

The preceding discussion applies to self-excited instabilities in
discs, but it has long been known that bi-symmetric spirals can also
be driven by bars \citep{SH76, vAR81, Li15} or tidal perturbations
\citep{TT72, No87, To81, BH92}, for which there is also observational
support \citep{KN79, KKC11}.  As reviewed in the introduction, there
is some evidence for a weak bar in M33 with a semi-major axis that is
perhaps only 360~pc \citep{RV94} which, if it is there at all, is
associated with the inner ends of the much more extensive spiral arms.
If the bi-symmetric spiral pattern observed in M33 is a response to
external forcing, then it weakens the above conclusion that its
bi-symmetry is an indicator of a heavy disc.  We here report
additional simulations to determine whether the pattern could be the
response to either type of forcing.

\subsubsection{Spiral response forced by an inner bar}
Our objective here is to test whether the large-scale bi-symmetric
spiral pattern in the old stellar disc of M33 could be the forced
response to the very mild inner bar.  We therefore report additional
simulations of the low-mass disc model in which we apply an external
forcing field to represent the bar.

The amplitude of the imposed potential varies as
\begin{equation}
\Phi_{\rm bar}(r,\theta,\phi,t) = \Phi_b(r,\theta,\phi,t) 
\left[ 1 + {1\over2}{\rm tanh}\left({t - t_0\over \tau_b}\right)\right],
\label{eq.barf}
\end{equation}
causing it to rise from near zero to its full value over the period of
$-2\tau_b \la \hbox{$t - t_0$} \la 2\tau_b$.  We adopt the bar-like
quadrupole form that rotates at the angular rate $\Omega_p$
\begin{equation}
\Phi_{\rm b}(r,\theta,\phi,t) = -{GM_{\rm b} \over a_1^3} {\alpha_2 r^2
\over 1 + (r/a_1\beta_2)^5}\sin^2\theta e^{2i(\phi-\Omega_pt)}.
\label{eq.barquad}
\end{equation}
The semi-major axis of the bar is $a_1$, $M_{\rm b}$ is the mass of the
assumed uniformly-dense ellipsoid, and the dimensionless constants
$\alpha_2$ and $\beta_2$ depend on the axis ratios of the bar $a_1:a_2$
and $a_1:a_3$ and are chosen to give the correct asymptotic variation of
the quadrupole potential at small and large radii \citep[see][]{We85,
  Se08}.

\begin{figure*}
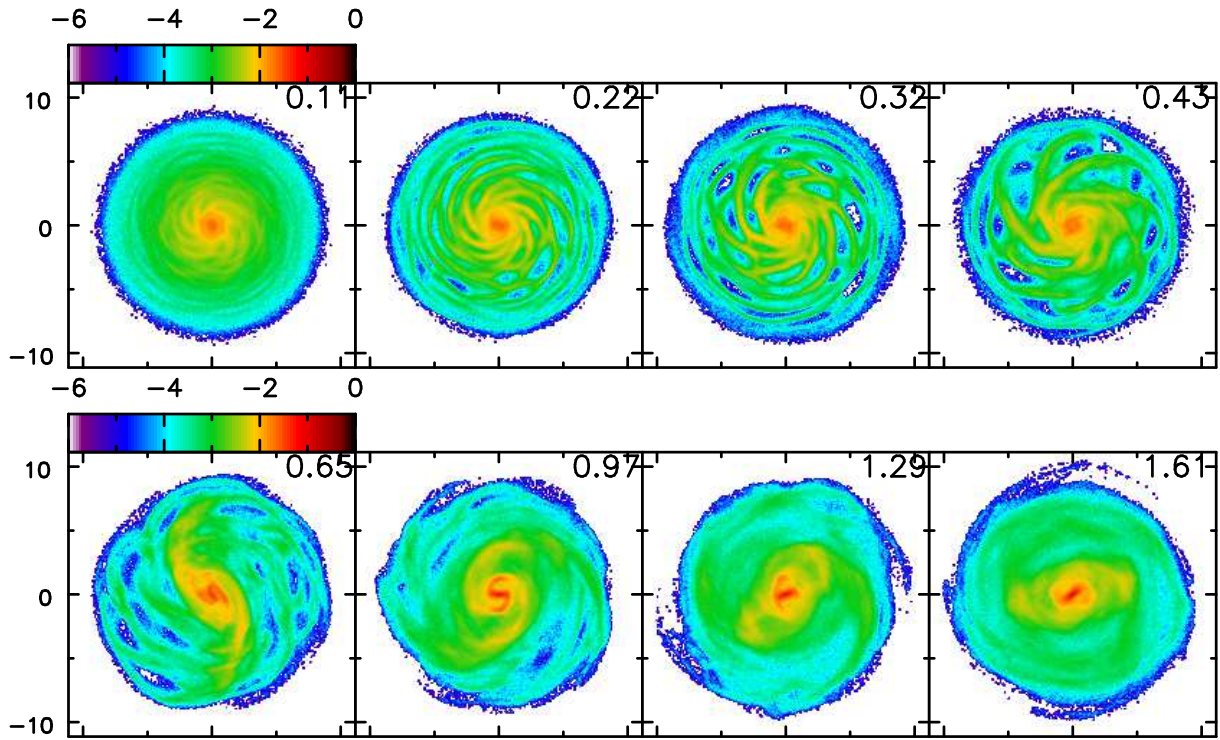

\includegraphics[width=.9\hsize,angle=0]{barfrc.ps}
\includegraphics[width=.9\hsize,angle=0]{tidfrc.ps}
% /home/sellwood/M33/barfrc.s /home/sellwood/M33/tidfrc.s
\caption{The evolution of the stellar component when the half-mass
  stellar disc is forced by a small bar (top row) and an externally
  imposed potential to mimic a tidal perturbation (bottom row).  The
  details of the externally applied fields are given in the text.  As
  in Fig.~\ref{fig.denevol}, times in the top right corner of each
  panel are in Gyr, the axes are marked in kpc, and the colour scale
  shows the logarithm relative to the maximum in each panel.}
\label{fig.forced}
\end{figure*}

We deliberately exaggerated the size and mass of the bar in M33, since
a smaller, weaker bar would have less effect.  Accordingly we set the
semi-major axis of the bar $a_1=0.4\;$kpc, made it a prolate homogeneous
ellipsoid with axes $a_1/a_2 = a_1/a_3 = 2$ that rotates about a short axis,
and adopt $M_{\rm b} = 3 \times 10^8\;$\Msun.  We hold $\Omega_p =
V_c(R_c)/R_c$ constant such that corotation is at $R_c = 1.2a_1$, and
choose the bar turn-on time scale $\tau_b = 2\pi/\Omega_p$.

The early part of the evolution of the stellar disc in the resulting
simulation is shown in the top row of Fig.~\ref{fig.forced}.  The
imposed potential elicited a supporting response from the particles in
the very center of the disc, but the evolution of the outer disc at
all times is barely distinguishable from that in Fig.~\ref{fig.light}.
This similarity is hardly surprising given the rapid outward decay of
the perturbing field (eq.~\ref{eq.barquad}).  We have not examined the
gaseous response, since we wish to know how the old stellar disc would
respond, but previous work \citep[\eg][and references therein]{Li15}
has shown that the spiral response of the gas is transient as the bar
turns on and extends only as far as the outer Lindblad resonance,
which is at $R \simeq 1.7\;$kpc for our adopted pattern speed.  Thus
we find that the large-scale bisymmetric spiral in the old stellar
disc of M33 could not be the forced response from any visible bar in
that galaxy.

\subsubsection{Spiral response forced by a tidal encounter}
Rather than attempt fully self-consistent simulations of encounters
between two galaxies, such as reported by \citet{SLSA}, we here follow
the simpler approach pioneered by \citet{To81}.  When two galaxies
move on eccentric orbits about their common center of mass, the
internal tidal stresses within each can be approximated as a rotating
quadrupole perturbation that is strong during the period of
peri-centre passage only.  Our objective is not to match any specific
perturber or orbit, but merely to explore how the disc responds to a
generic stretch of this kind.

In this case, we adopt the following time dependence for the
perturbing potential
\begin{equation}
\Phi_{\rm tide}(r,\theta,\phi,t) = \Phi_b(r,\theta,\phi )
{e^{-(t-t_0)^2/2\tau_t^2} \over \tau_t \sqrt{2\pi}},
\label{eq.tide}
\end{equation}
where $\tau_t$ is the approximate duration of closest approach, and we
adopt the same form for $\Phi_b$ as given in eq.~(\ref{eq.barquad}).
In order to achieve the desired duration, we choose $\tau_t = 1 /
\Omega_p$, causing the perturbing field to be strong as it turns
through an angle of $\sim 2\;$radians.  We choose $a_1=20\;$kpc, set
$\Omega_p = V_c/a_1$, with $V_c = 120\;$km~s$^{-1}$, $t_0 = 2.5\tau_t$
so that the perturbation peaked $t \simeq 0.42\;$Gyr after the start
of the simulation.  This somewhat arbitrary set of values deserves
some discussion.  The quadrupole field (eq.~\ref{eq.barquad}) rises
simply as $r^2$ for $r \ll a_1$, so that other choices for $a_1 \gg
10\;$kpc would merely rescale the perturbation amplitude within the
disc.  We chose a prograde, in-plane encounter since it would excite
the greatest response, and moderately inclined orbits should exert
similar forces in the disc plane.  Our choice of a constant $\Omega_p$
matters only while $|t-t_0| \la \tau_t$, since the forcing frequency
is immaterial when the perturbation is weak.  The circular orbit
frequency at $R=a_1$ corresponds to a higher-than-circular frequency
during a peri-centric passage at a larger radius, although the
frequency would be lower if the orbit were also inclined.  Thus the
only significant free parameter is the perturbation amplitude.  Here
we choose $M_{\rm b}$ such that the peak perturbed acceleration at the
disc edge ($R=10\;$kpc) is 2.5\% of the centripetal acceleration at
the same point.

The evolution after the closest passage is shown in the bottom row of
Fig.~\ref{fig.forced}.  In this case, the external field has indeed
provoked an initial bi-symmetric spiral response in the stellar disc,
which is tightly wrapped in the very centre, while the outer disc
supports an oval distortion.  The oval distortion persists at
later times, while the stronger spiral features fade, leaving a
double-bar structure.  An additional simulation with twice the
amplitude of the perturbing field quickly formed a strong bar.

The obvious candidate for a tidal perturbation of M33 is a possible
encounter with M31, and the models of \citet{SLSA} find a best fit for
the warp when closest approach occurred about 1~Gyr ago.  However, the
inner disc of M33 at the last two frames in the bottom row of
Fig.~\ref{fig.forced} that bracket 1~Gyr after the closest approach
has a clear double bar structure, which is quite inconsistent with the
morphology of M33 now.  As there is also the argument that M33 may be
on its first approach to M31 \citep{vdM18}, we do not consider forcing
by M31 an attractive explanation for the current spiral appearance of
M33.  If the spirals in our low-mass disc model of M33 are to have
been tidally induced, they must have been forced by the very recent
passage of a substantial unseen companion.

\Ignore{Even so, the slowly evolving tightly-wrapped inner spiral most
  clearly visible in Fig.~\ref{fig.forced} (bottom row, second frame,
  $t \sim 0.5\;$Gyr after closest approach), is not seen in the
  near-IR images of M33.}

\subsubsection{Conclusion on disc mass}
Halving the disc mass from the value in our fiducial model did allow
the disc to remain stable.  The above discussion has not only stressed
that the stellar M/L of this stable model is extraordinarily low, but
the self-excited spirals in the disc are multi-armed, which is
inconsistent with the pattern observed in the old stellar disc.  We
have shown that the observed spiral could not be the forced response
to any reasonable visible, bar in the inner disc of M33.  We have
found that a transient, large-scale bi-symmetric spiral can be excited
in a stable low-mass disc by a recent tidal encounter, although we
argue that M31 seems a most unlikely culprit.  Without extensive
additional exploration, we cannot exclude the doubly speculative
scenario that the disc of M33 has an extremely low M/L and the
bi-symmetric spiral results from of a favourable tidal perturbation by
a substantial unseen companion within the last $0.5\;$Gyr.

Albeit with this caveat, it seems unlikely that the solution to the
stability of M33 is that the disc has a lower-than-expected mass.

\subsection{Stability and random motion}
We successfully inhibited bar formation in our simulation by setting
$Q = 2$ and doubling the disc thickness, as shown by the green line in
Fig.~\ref{fig.ampevol}.  The projected velocity dispersion of the star
particles near the disc centre in this case is $\sim 43\;$km~s$^{-1}$,
which is substantially larger than the observed dispersion of 25 --
35~km~s$^{-1}$.  Note that it is possible that the measurements of the
stellar velocity dispersion could be biased low \citep[\eg][]{An16}
because the brighter stars belong to a younger population having less
random motion, and therefore the mass-weighted velocity spread in the
disc of M33 could be larger than the reported luminosity-weighted
values.  Note that \citet{KM93} were aware of this possible bias, and
deliberately selected the Ca II triplet line for their measurement, as
did \citet{CW07}, because a line in the red part of the spectrum would
yield a value more representative of the older stars than would other
lines.

Somewhat surprisingly, the stellar disc in this model did manifest
some mild multi-arm spiral activity, which we attribute to responses
to instabilities in the gas component.  The surface mass density of
the gas disc is substantially lower than that of the stars, causing it
to prefer instabilities on a small spatial scales, as discussed above,
and these disturbances seemed able to elicit a weak suporting response
from the stars.  However, the amplitudes of two-armed disturbances in
the combined star-gas disc remained very low throughout
(Fig.~\ref{fig.ampevol}), so that this hot-stellar disc model was
unable to reproduce the dominant $m=2$ spiral pattern in near-IR
images of M33.

Thickening our disc model slightly reduces the innermost gradient of
the circular speed that arises from the disc component, and allows the
central halo density to be slightly larger.  However, disc thickness
must be supported by vertical velocity dispersion.  As we discussed
above, our models really should satisfy the observational constraint
on the line-of-sight velocity dispersion, and therefore invoking
increased vertical dispersion would require us to reduce the in-plane
random velocities, although we ignored this constraint in the above
test.  Were we to take account ot this constraint while increasing the
disc thickness to allow a slightly higher central halo density, we
really should also have reduced the in-plane motions, which would have
been destabilizing.

\citet{SM94} demonstrated that heavy discs could be globally stable
when they contain two almost equal cool components of direct and
counter-rotating stars, such as was found in NCC 4550 \citep{Ru92,
  Ri92}.  Unfortunately, the low surface brightness of the M33 disc
creates challenges to measurements of the stellar velocity
distribution and absorption-line measurements are available only very
close to the nucleus of this galaxy \citep{KM93, CW07, Ko10}.  We
therefore have little direct evidence against a possible large
counter-rotating stellar population in M33.  However, the existence of
a regular spiral pattern seen in the NIR images is good indirect
evidence that the stellar disk has significant net rotation in the
same sense as the gas disc, and that counter-rotation is an unlikely
explanation for the absence of a bar within the disc of M33.

\subsection{More exotic ideas for stability}
\citet{Mi83, Mi15} has argued that observed orbital speeds in galaxies
that are higher than expected from the Newtonian attraction of the
visible matter need not imply dark matter if the laws of Newtonian
dynamics break down at weak accelerations.  His hypothesis, known as
MOND, introduces a critical acceleration scale with a value $a_0
\approx 10^{-8}\;$cm~s$^{-2}$ at which gravitational attraction
transitions from the Newtonian behaviour to a slower decrease, and it
has met with surprising success \citep{SM02}.  (The radial
acceleration at $R=8\;$kpc in M33 is $V^2/R \approx 0.5 \times
10^{-8}\;$cm~s$^{-2}$.)  \citet{BM99} found that as the mean
acceleration in the disc was lowered from the Newtonian to the MOND
regime the instability to bar formation was reduced, until the degree
of stability levelled off deep in the MOND regime.  Other simulations
by \citet{TC07} and \citet{LFK} confirmed that bars still form in
MOND, although the final bars are shorter and weaker than in
comparable Newtonian simulations with responsive halos because bar
secular growth is suppressed.\footnote{I. Banik (2019, private
  communication) confirmed that a model of M33 computed using a MOND
  gravity law did form a strong bar.}  A different non-Newtonian
gravity law, known as MOG \citep{MR13}, invokes a Yukawa-like law of
gravitational attraction, and \citet{Ro18} also found that the bar
instability was weakened, but not suppressed, in simulations that
employed this law.  Thus neither of these hypotheses, seems able to
account for the apparent stability of M33.

A rigid, unresponsive halo would be unphysical for most conventional
dark matter candidates.  However, axion-like particles, sometimes
described as fuzzy dark matter (FDM), are increasingly being discussed
as a possible dark matter candidate for a number of reasons
\citep[\eg][]{Hu17}.  Halos composed of such ultra-light, quantum
mechanical, bosonic particles are believed to form a quasi-uniform
density condensate \citep{MP15}, known as a ``soliton'' core, which
transitions to a declining density outside the core.  It is unknown
whether such a halo core would behave more like a rigid mass
component, but its response to rotating non-axisymmetric distortions
in the disc might be expected to be very different from that of usual
collisionless dark matter.

However, the soliton core is expected to be small, and the dynamical
behaviour of FDM outside the core is the same as for any other
collisionless DM candidate.  The core radius is of order the de
Broglie wavelength of the particles, and the core mass for a given
mass halo is also set by the particle mass and fundamental constants.
Using the virial mass of $M_{\rm vir} \sim 5 \times
10^{11}\;$M$_{\odot}$ for M33 \citep{Ka17} in equations (29 -- 31) of
\citet[][see also \citet{Sc14a}]{Hu17}, together with their suggested
mass for the FDM particle of $m_{\rm FDM} \simeq 10^{-22}\;$eV, we
find a soliton core mass $M \sim 10^{9} M_{\odot}$ with the half-mass
radius $r_{1/2} \sim 335\;$pc.  These values imply a central density
for the soliton core of $\rho_{\rm c} \sim 7 \;$\Msun\ pc$^{-3}$, over
30 times higher than our fitted value (eq.~\ref{eq.haldens}), which
would cause the total rotation curve to rise more steeply than is
observed (Fig.~\ref{fig.vcomps}).  Since the value of $m_{\rm FDM}
\simeq 10^{-22}\;$eV \citep[\eg][]{SCB14} should be universal, the
fact that the expected soliton core in M33 is too dense could present
a difficulty for FDM models.  Setting that concern aside, it is clear
than the dynamics of an FDM halo should be similar to that of any
other collisionless DM candidate at radii $r \ga 0.5\;$kpc while the
bar that forms in the disc is several times larger.  Thus a soliton
core does not seem like a promising solution to the challenge
presented by the apparent stability of M33.

\section{Conclusions}
\label{sec.conclusions}
The inner disc of the Local Group galaxy M33 appears to be in settled
rotational balance and the old stellar distribution, as revealed in
near IR images, supports a mild, bi-symmetric spiral pattern with no
sign of a strong bar.  The origin of the warp in the outer disc is
uncertain, but even if it had been tidally induced during an encounter
with M31 some 1~Gyr ago \citep{SLSA}, the inner disc within $\sim
8\;$kpc of the centre appears to have been little affected by the
disturbance.

We have constructed a good equilibrium model of the inner parts of M33
that matches the surface mass density of the stars and gas with a dark
matter halo that fits the well-observed rotation curve of the galaxy.
The stellar velocity dispersion in our model matches the central
value, which is the only observational constraint, and we assume $Q =
1.5$ and a constant disc thickness at all radii.

We have computed the evolution of this model, which carefully matches
all available observational constraints, in order to determine whether
it is stable.  We find that it is not, and it forms a strong and
persisting bar with a semi-major axis of between $2 \la a_B \la
3\;$kpc on a time scale $<1\;$Gyr.  We discount the possibility that
the properties of the inner disc of M33 have been changed within the
last Gyr, or 2.5 orbit periods, in order to cause it to be unstable
today when it was stable earlier.

Further simulations have shown that the instability persists no matter
what prescription we use to model the gas.  Raising the mass of the
nuclear star cluster or making the halo counter rotate in the opposite
sense of the disc did change the evolution, but bars still formed and
the final states of the simulations were inconsistent with the current
observed state of the galaxy.

We succeeded in preventing a bar from forming at all in two models,
and found a substantially weakened the instability in a third.
Halving the stellar disc mass prevented a bar from forming entirely,
but the mass-to-light ratio in solar units of the stellar disc was
changed to $\Upsilon_V \sim 0.6$ and $\Upsilon_K \sim 0.23$ from the
lower value preferred by \citet{Co14}, and such values are well below
the ranges that are believed to apply in most galaxies \citep{Co13,
  SML18}.  Increasing the random motion of the disc stars also
suppressed the bar, but the random motions exceeded those observed
\citep{KM93, CW07}.  Replacing the responsive dark matter halo by the
fixed potential of a mass distribution of the same density resulted in
a shorter and weaker bar.  However, all three of these ``solutions''
are unattractive for a different reason also: the self-excited spiral
patterns that developed in the stellar component of the simulations
were multi-armed, and were quite inconsistent with the lare-scale
bi-symmetric pattern seen clearly in images taken in the near IR.  We
did obtain a transient bi-symmetric spiral response to a simplified
tidal forcing of a low-mass disc model, but to argue that the current
state of M33 can be explained in this way requires both an
exceptionally low M/L for the stellar disc and the recent passage of a
substantial unseen companion on a favourable orbit.

We discussed, but did not test, modified gravity laws and soliton
cores of ``fuzzy dark matter'', concluding that neither seemed to
offer a promising solution to the puzzle presented by apparent
stability of the inner disc of M33.  We therefore conclude that the
survival of the currently observed state M33 is not yet understood.

We selected M33 for this study because it is so well-observed and
enables us to make the clearest possible case, but we strongly suspect
that many other galaxies would present a similar challenge.  It is
shocking that we still do not understand how the bar instability is
avoided in real galaxies, despite \citet{Ho71} having first sounded
the alarm about its prevalence nearly half a century ago.  Our direct
tests have shown that none of the ideas proposed over the subsequent
years can account for the absence of a bar in M33.  The lack of a
satisfactory explanation is perhaps central to another outstanding
issue, also noted in the opening paragraph of the introduction, of
what determines the distribution of bar strengths in galaxies?  These
stubborn problems present still daunting challenges to galactic
dynamics.

\section*{Acknowledgements}
We thank John Kormendy, Claude Carignan, Hsi-Yu Schive, Justin Read,
Larry Widrow, and the referee, Paolo Salucci, for comments on a draft
of the paper.  This work was begun during a visit by JAS to Shanghai
Observatory, which was supported by Grant No.\ 2018VMA0051 from the
Chinese Academy of Sciences President’s International Fellowship
Initiative, and he also acknowledges the continuing hospitality of
Steward Observatory.  The paper was completed at KITP, which is
supported in part by NSF grant PHY-1748958.  The research presented
here was partially supported by the National Key R\&D Program of China
under grant no.\ 2018YFA0404501, by the National Natural Science
Foundation of China under grant nos.\ 11773052, 11761131016, 11333003,
and by a China-Chile joint grant from CASSACA. JS acknowledges the
support of a {\it Newton Advanced Fellowship} awarded by the Royal
Society and the Newton Fund.  This work made use of the facilities of
the Center for High Performance Computing at Shanghai Astronomical
Observatory.


% Don't change these lines
\bsp	% typesetting comment
\label{lastpage}
\end{document}